\documentclass[11pt, onecolumn]{IEEEtran}

\usepackage{times}
\usepackage{epsf}
\usepackage{latexsym}
\usepackage{amsmath}
\usepackage{amsfonts}
\usepackage{euscript}
\usepackage{url}

\newcommand{\ignore}[1]{}

\newcommand{\isequal}{\stackrel{\rm ?}{=}}

\newcommand{\qed}{\hspace*{\fill}\rule{7pt}{7pt}}

\newenvironment{Proof}{\medskip \noindent{\bf Proof.}\hspace*{0.2cm}}{\qed\medskip}

\newcounter{defcounter}
\newtheorem{Thm}{\noindent \hskip -0.5cm {\bf Theorem}}
\newenvironment{Def}{\medskip \noindent {\bf Definition \thedefcounter.}}{\hspace*{\fill}\stepcounter{defcounter}\medskip}

\font\thirdteen=cmr10 at 13pt

\font\sixteen=cmr10 at 16pt

\begin{document}

\title{\sixteen{On the Compression of Cryptographic Keys}}
\author{\thirdteen{Aldar C-F. Chan\\National University of Singapore}}

\maketitle

\begin{abstract}
Any secured system can be modeled as a capability-based access
control system in which each user is given a set of secret keys of
the resources he is granted access to. In some large systems with
resource-constrained devices, such as sensor networks and RFID
systems, the design is sensitive to memory or key storage cost. With
a goal to minimize the maximum users' key storage, key compression
based on key linking, that is, deriving one key from another without
compromising security, is studied. A lower bound on key storage
needed for a general access structure with key derivation is
derived.  This bound demonstrates the theoretic limit of any systems
which do not trade off security and can be treated as a negative
result to provide ground for designs with security tradeoff. A
concrete, provably secure key linking scheme based on pseudorandom
functions is given. Using the key linking framework, a number of key
pre-distribution schemes in the literature are analyzed.
\end{abstract}

\section{\label{intro} Introduction}
In any computer system offering security-related services, it is a
basic necessity that its users have access to some private
information to give them leverage over an adversary. These secret
pieces of information are commonly known as (cryptographic) keys.
The key is usually used as input to protocols or algorithms for
identification, secrecy and authentication purposes.  Nearly all
such systems can be modeled as a capability-based access system in
which each resource is assigned a secret key and a user granted with
access right to the resource would be given its key. For example, in
secure group communication \cite{blun92}, each conference group is
assigned a conference key which is given to all users belonging to
the group so that the communication of the group could be kept
secret and message authentication can be achieved within the group.

Ideally, the security requirement of a typical system (not limited
to secure group communication) is that all users outside a
particular group or not granted access to a resource should not be
able to obtain or compute the key for it even by collusion.  For
instance, in secure group communication, it is necessary to ensure
that all users outside a certain conference group (whose key is
treated as a resource key) should not be able to derive the group
key from their keys.

In most cases, the storage needed at each user could be too large to
be practical.  For example, in a typical access control system, if a
user has a high level of privilege, his device may need to store a
considerable number of keys.  Since the cost of the tamper-resistant
storage for the keys increases linearly with the size of the key
storage, it is thus worthwhile to study techniques to generate all
these keys from a smaller seed or compress the key materials.  There
is a similar problem facing emerging applications like sensor
networks and RFID tags.  Despite the dropping cost of secure
storage, key storage is still a big concern in these applications,
involving low cost embedded devices which have to store a
considerable amount of secret keys. Compressing key materials is
essential to the scalability of such designs.

To ensure correctness of the operation of all cryptographic
algorithms, the key compression needs to be lossless.  Besides, to
protect a resource from unauthorized access by collusion of
compromised users, the key compression should not leak information
that can ease unauthorized access to any resource key not given to
the compromised users.  This paper studies techniques to create
dependency between resource keys (to derive one key from another) so
as to reduce the storage requirement on each user device.  In other
words, we exploit the redundancy in privileged group memberships for
key compression.  The goal is to minimize the maximum of user key
storage over all users. To link keys together, we need to consider
the access memberships of all the resources in the system to avoid
compromising the security of some resources.\footnote{The access
membership of a resource is the subset of legitimate users granted
access right to it.}  We investigate the limit of this key
derivation approach by deriving a bound on maximum compression
achievable without compromising the security of any resource key.

Due to their simplicity, existing work in the literature such as
\cite{DFM06,WC01,Chien04} only considers monotonic access
structures. Whereas, this paper considers a much more general access
structure without posing any restrictions on what properties it must
have. The results of this paper are general enough to cover most
practical application scenarios. Note also that the applicability of
the model we use is not limited to symmetric or shared key systems.
For asymmetric key systems, the model depicts the possession of
private keys and the resources represent all algorithms requiring a
private key input. For instance, a resource could represent the
decryption algorithm of a certain public key cryptosystem and its
keys represent the required private keys to achieve a successful
decryption of a certain ciphertext encrypted using the corresponding
public key.  The access control model we consider in this paper
could cover a wide range of actual systems, including those not
designed for access control purposes.

The contribution of this paper is three-fold. First, we derive the
lower bound on key storage needed for a general access structure if
key dependency is created between keys held by a user. This lower
bound corresponds to the theoretical limit on maximum key
compression achievable in an ideal access structure without key
compromise. We also show that this bound is tight by giving some
concrete examples in sensor network key pre-distribution, which are
either bound-achieving or close to this bound. Second, we give a
practical, provably secure key linking scheme (for a general access
structure) based on pseudorandom functions (PRF). We also provide a
reduction proof of security for this construction.  Third, we
demonstrate how to apply the key linking framework to reduce key
storage in pairwise key pre-distribution schemes for sensor
networks.  We have to emphasize that, unlike the existing schemes
with key re-use such as \cite{esch02,CPS03}, the resulting key
storage reduction does not come with a price of lowering the
resilience or security against compromised nodes.\footnote{In nearly
all of the existing key pre-distribution schemes for sensor
networks, in order to lower the key storage requirement, the same
key is used for links between several pairs of nodes.  So when a key
is exposed to an adversary due to a compromised node, all these
links will be compromised instead of one, thus lowering the
resilience of the network against compromised nodes.} The only
trade-off is lowering the security guarantee from the
information-theoretic sense to the
computational-complexity-theoretic sense (due to the use of
pseudorandom functions), which in essence makes no difference in
practice.

In the next section, we present the definitions of access
structures.  In Sections \ref{bound} and \ref{pseudo}, we present
the key storage lower bound and the key linking construction based
on pseudorandom functions respectively. Then we consider applying
the key linking framework to key pre-distribution for sensor
networks in Section \ref{ks-sensor}. Finally, we have some
discussions in Section \ref{discuss} and conclude in Section
\ref{conclude}.

\section{\label{access} Access Structure}

\begin{figure}[hbt]
 \vskip -2cm
    \centerline{
        \epsfysize=18cm
        \epsffile{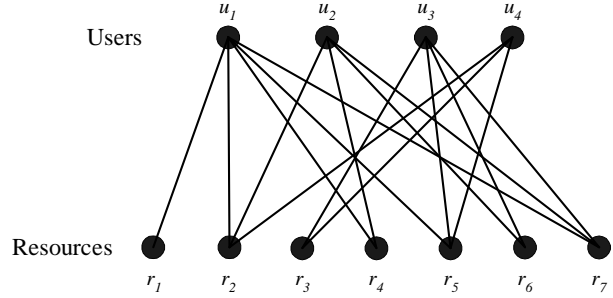}\\
        }
        \vskip -12.5cm
        \caption{\label{ks-access-structure} A Typical Access Structure Graph}
\end{figure}

We use an access control system to model a security system; a fairly
wide range of applications can be covered by this model. The access
structure of a typical system depicts the relations between users
and keys/resources.  A graphical presentation is shown in Figure
\ref{ks-access-structure}.

Suppose ${\cal U} = \{u_1, u_2, \ldots, u_n\}$ is the set of users
and ${\cal R} = \{r_1, r_2, \ldots, r_m \}$ the set of resources in
a system.  Let $2^{\cal U}$ be the set of all subsets of ${\cal U}$
and denote the set of all possible secret keys by ${\cal K}$.  Each
resource $r_j \in {\cal R}$ is associated with a key $k_j \in {\cal
K}$ and an ordered pair $(P_j, F_j)$ with $P_j \subseteq {\cal U}$
and $F_j \subset 2^{\cal U}$; $P_j$ is the subset of privileged
users granted access to $r_j$ whereas each element in $F_j$
corresponds to a forbidden subset of users which should not be able
to access $r_j$ even if all of them collude. Then the access
structure of a system has the following definition.

\begin{Def}
The access structure $\Gamma$ of a security system $({\cal U}, {\cal
R}, {\cal K})$ is the following set of 4-tuples: $\{(r_j, k_j, P_j,
F_j): r_j \in {\cal R}, k_j \in {\cal K}, P_j \subseteq {\cal U},
F_j \subset 2^{\cal U}\}$.
\end{Def}

In the definition of an access structure, a system is not required
to guard against all illegitimate users outside the privileged group
of a resource from accessing it.  In practical scenarios, usually,
only a bounded number of illegitimate users in collusion could be
excluded; there is a tradeoff of security for storage. However, this
paper considers an ideal access structure which is the most desired
setting as raised by Naor et. al. \cite{naor01} in the context of
broadcast encryption. An access structure is ideal if all the
illegitimate users to any resource in the system are excluded from
accessing it.

\begin{Def}
An access structure $\Gamma  = \{(r_j, k_j, P_j, F_j) \}$ for a
security system $({\cal U}, {\cal R}, {\cal K})$ is ideal if ${\cal
U} \backslash P_j \in F_j$, $\forall r_j \in {\cal R}$.
\end{Def}

In a security system, the access structure is associated with a key
assignment scheme. The set of keys held by a user may not be exactly
the same as that of the resources he could access, but should allow
him to compute all the resource keys he needs.  An access structure
graph, whose definition is given below, incorporates a key
assignment to an access structure.

\begin{Def}
Given a set of users ${\cal U} = \{u_1, u_2, \ldots, u_n \}$, a set
of resources ${\cal R} = \{r_1, r_2, \ldots, r_m \}$ and a set of
keys ${\cal K}$, an access structure graph ${\cal G}$ for the system
is a bipartite graph with vertex set $V({\cal G}) = {\cal U} \cup
{\cal R}$ and edge set $E({\cal G}) \subseteq {\cal U} \times {\cal
R}$, and the following properties hold:
\begin{itemize}
\item
$(u_i, r_j) \in E({\cal G})$ if and only if $u_i$ can access $r_j$.

\item
Each resource vertex $r_j$ is associated with a privileged user
subset $P_j \subseteq {\cal U}$ such that $(u_i, r_j) \in E({\cal
G})$ if and only if $u_i \in P_j$.

\item
Each resource vertex $r_j$ is associated with a key $k_j \in {\cal
K}$.
\end{itemize}
\end{Def}

The associated key assignment of an access structure graph is said
to be secure and sound if the following holds: a user $u_i$ can
compute the key $k_j$ if and only if $(u_i, r_j) \in E({\cal G})$
for all $1 \le j \le m$. Note that existing key pre-distribution
schemes for sensor networks with key re-use \cite{esch02, CPS03} do
not satisfy this requirement of security and soundness.

\section{\label{bound} A Key Storage Lower Bound with Key Dependency}
This section uses the access structure graph defined in Section
\ref{access} to derive a lower bound on the key storage requirement
if dependency is created between keys.

In an access structure graph, the degree of each user vertex $u_i$
is the key storage requirement at $u_i$ assuming the users store the
resource keys directly and each key has the same length, whereas,
the degree of each resource vertex $r_j$ is the number of privileged
users who can access it, which is the same as $|P_j|$.  Let the key
storage at user $u_i$ be $d_i$, the goal is to minimize $\max_{u_i
\in {\cal U}}\{d_i\}$.

Usually the resource keys should be picked independently at random
to ensure security. However, for some users, storing multiple keys
may be redundant.  For instance, if a privileged group $P_1$ is the
subset of another say $P_2$, that is, $P_1 \subset P_2$, then it is
redundant for a user in $P_1$ to store $k_2$ (the key for $P_2$) in
addition to $k_1$ (the key for $P_1$). If $k_2$ can be derived from
$k_1$, then the storage at each $u_i \in P_1$ would be reduced by
one key\footnote{Such a derivation is possible if $k_2$ would not
leak out information about $k_1$ practically.  We show in the next
section how such a derviation can be instantiated by a pseudorandom
function.}, equivalent to removing the edge $(u_i, r_2)$ from ${\cal
G}$ and adding a new edge between $r_1$ and $r_2$ (the key
dependency). Note that the resulting graph is no longer bipartite.

Given two keys $k_j$ and $k_j'$ for privileged subsets $P_j$ and
$P_j'$, if $k_j'$ is derived from $k_j$, all users in $P_j$ would
have access to $k_j'$. As a result, to ensure that the key linking
does not compromise security, it is necessary to make sure that $P_j
\backslash P_j' = \phi$ (the empty set). In other words, $P_j
\subset P_j'$ if $P_j' \neq P_j$. Otherwise, a user not in $P_j'$
(but in $P_j$) would have access to $k_j'$.  Subject to this
constraint, the best achievable key storage reduction is given by
the following theorem.

\medskip
\begin{Thm}\label{ch2-thm-hash}
If dependency is created between keys while maintaining the ideal
access structure and security of a system, depending on the access
structure, the best achievable maximum storage at each user is at
least $\lceil \frac{m}{n} \rceil$ where $n$ is the total number of
users and $m$ is the total number of resources with distinct access
membership.
\end{Thm}
\begin{Proof}
To maintain the security and access structure, a key $k_j'$ can be
derived from another key $k_j$ only if $P_j \subset P_j'$ and the
users in $P_j' \backslash P_j$ need to store $k_j'$ while users in
$P_j$ can generate $k_j'$ from $k_j$.

If a key $k_j'$ can be generated from $k_j$, then $|P_j' \backslash
P_j| \ge 1$ since $P_j \subset P_j'$ (Note that the $m$ resources
have distinct access membership).  That is, at least one user in
$P_j'$ needs to store $k_j'$. In other words, after key linking,
each resource vertex in the access structure graph should have at
least one edge coming from the set of user vertices. If we denote
the number of edges coming from a user vertex to $r_j$ by $y_j \ge
1$ and the degree of a user $u_i$ by $x_i$, then
\[
\sum_{i=1}^n x_i = \sum_{j=1}^m y_j \ge \sum_{j=1}^m 1 = m.
\]
In the best case, the degrees of any two users $u_i$ and $u_i'$
should not differ by more than 1. Hence, the maximum user degree
$\max_{u_i \in {\cal U}} deg(u_i) \ge \lceil \frac{m}{n} \rceil$.
\end{Proof}

The result of Theorem \ref{ch2-thm-hash} does not assume any
concrete construction for creating the key dependency. It is rather
general, discussing whether a particular key could be derived from
another while maintaining security and what the best achievable key
storage reduction would be.  In the best scenario, a $\frac{1}{n}$
reduction factor could be achieved by eliminating all redundancy in
the privileged group memberships of a system. The lower bound in
Theorem \ref{ch2-thm-hash} is also tight as can be seen from the
example below.

\begin{figure*}[hbt]
    \vskip -1cm
    \centerline{
        \epsfysize=17cm
       \epsffile{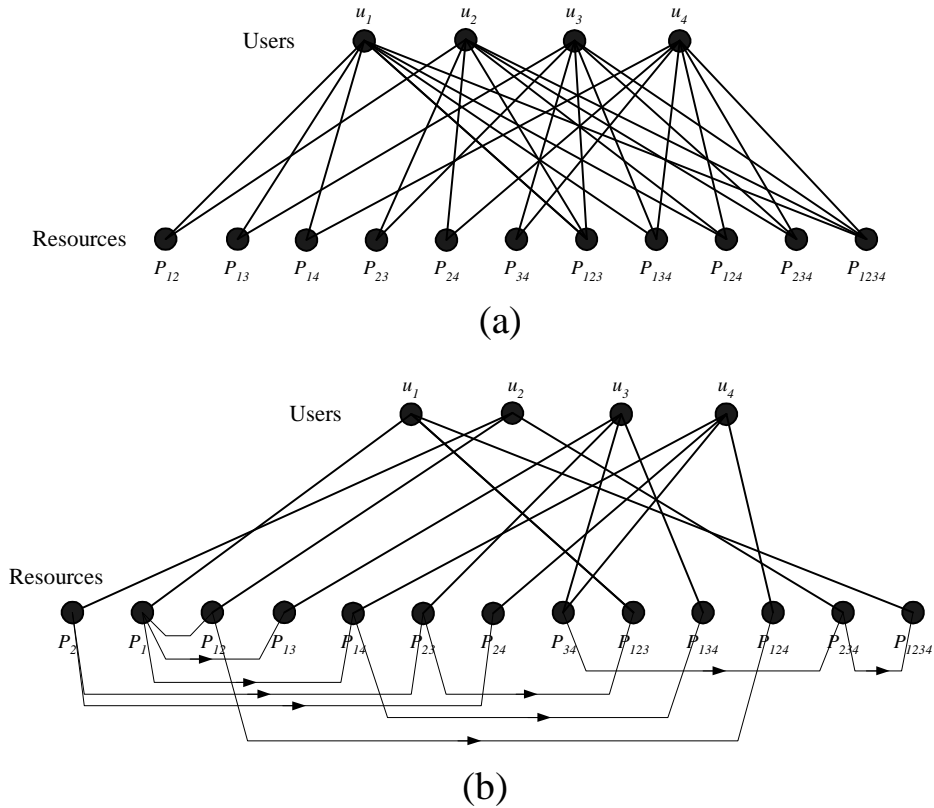}\\
        }
        \vskip -5cm
        \caption{\label{ks-access-structure-link} The Access Structure Graph of a Complete KPS Scheme (a) before Key Linking and (b) after Key Linking}
\end{figure*}

Shown in Figure \ref{ks-access-structure-link} is an example for the
complete secure group communication with 4 users.  Originally, each
user has to store $2^{4-1} - 1 = 7$ keys.  Note that $m=11$ and
$n=4$, and hence $\lceil \frac{11}{4} \rceil = 3$.  After key
linking\footnote{Note that two fictitious nodes are added to achieve
a lower storage; the lower bound stated in Theorem
\ref{ch2-thm-hash} holds here because the only effect of adding
these fictitious nodes is that two resource nodes are added to the
original access structure graph, which in essence increases the
value of $m$.}, the maximum number of keys of a user is $4 > 3$.

\begin{figure*}[hbt]
    \vskip -1cm
    \centerline{
        \epsfysize=17cm
       \epsffile{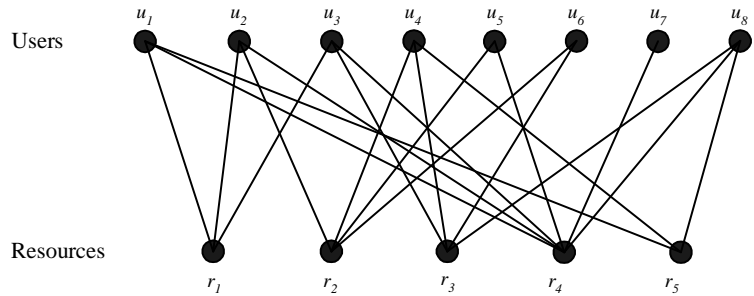}\\
        }
        \vskip -12cm
        \caption{\label{ks-access-structure-link-fail} An Example Access Structure Graph that Key Linking cannot lead to a Reduction on the Maximum Key Storage per User}
\end{figure*}

Clearly, as long as there exist resource nodes (in the access
structure graph) sharing a non-empty intersection between their
access membership sets, key linking could always be possible between
them, resulting in storage reduction at some users.  However, it is
not necessarily true that $\max_{u_i \in {\cal U}}\{d_i\}$, the
maximum of the key storage per user (over all users), can always be
reduced. We may achieve storage reduction at all users but the one
with maximum storage particularly when the access structure graph is
very irregular. For instance, shown in Figure
\ref{ks-access-structure-link-fail} is a case where key linking
cannot lead to a reduction on the maximum of key storage per user
(which is originally 3); no matter how key dependency is created,
the maximum key storage per user is still 3 while the lower bound
should be $\lceil \frac{5}{8}\rceil = 1$. Whether the lower bound on
the maximum key storage per user (as stated in Theorem
\ref{ch2-thm-hash}) can be achieved and whether key linking can
reduce the maximum key storage per user depends on the access
structure. Loosely speaking, if the access structure graph is dense,
it is likely that a reduction on the maximum key storage per user
can be achieved through key linking; if the degrees of user nodes
are regular (that is, each user has access to roughly the same
number of resources), and the sequence of the degrees of the
resource nodes (in ascending order) does not have a sharp difference
between two consecutive elements (that is, there does not exist a
resource node having a considerably larger access membership set
than others), key linking is most effective with the resulting
maximum key storage per user closest to the lower bound.

A general key linking algorithm is designed (given in the next
section) to run experiments on different access structure graphs.
The results actually agree with the above observations.

\section{An Algorithm to find a Key Linking Pattern}
\label{algo}

Depicted below is a general algorithm for finding a key linking
pattern for any given access structure graph. This algorithm
converts an access structure graph into one with key linking.  It
runs as follows:

\bigskip

  \begin{center}
      \begin{tabular}{p{5in}} \hline
      An algorithm to find a key linking pattern.\\
\hline
\\
Input: an access structure graph\\
Output: an access structure graph with key linking

\begin{enumerate}
\item Sort the resource nodes according to the size of their privileged
groups and assign an index (0,1,2, .....) to each node
accordingly.

\item Pick the node with largest index to start with and set it as
the current node.

\item From current node, pick a node with the next smaller index and set its index as the find-pointer.

\item Check if the node at find-pointer is a subset of the current node.
\begin{enumerate}
\item If yes, mark a link at the find-pointer node to the current node
and go to step 5.

\item If no, decrease the
find-pointer by 1 and repeat step 4 if the find-pointer is
greater than zero, otherwise, go to step 5.
\end{enumerate}

\item Decrease the current node index by 1.

\item Repeat 3-5 until the current node index is 0.

\end{enumerate}\\

\hline
\end{tabular}
\end{center}

\bigskip

\section{\label{pseudo} A Key Derivation Scheme based on Pseudorandom Functions}
In order to generate a key $k'$ from another key $k$, we could
consider $k$ as a seed to some pseudorandom generator $g$ which
outputs $k'$.\footnote{The output space of $g$ should be the same as
the key space.} The requirement of a suitable generator is
that,without the knowledge of the seed $k$, to any computationally
efficient algorithm (i.e. polynomial-time), the output of $g$ is
indistinguishable from any random number picked uniformly from the
key space.  This would ensure that the view to anyone
(computationally bounded and without the knowledge of the seed) is
almost identical to that without key linking, thus guaranteeing that
nobody could learn any information about the seed key from the
generated keys. This computational indistinguishability requirement
is essential to ensuring the security of the whole system.  The
explanation is as follows: Note that the resulting keys from the
generator is to be used as an input key to a certain cryptographic
algorithm or protocol whose security guarantee is usually based on
the assumption that the input key is uniformly picked from the key
space.  In fact, it can be shown that, if the distribution of the
key generator output is computationally
indistinguishable\footnote{Two probability distributions are said to
be computationally indistinguishable when no polynomial-time
distinguishing procedure can tell them apart.  In other words, given
a sample which could be picked from either of the two distributions,
no sufficiently efficient algorithm can tell whether the sample is
from the first distribution or the second.} from a uniform
distribution over the key space, the security guarantee of
cryptographic primitives like encryption and message authentication
codes holds.

Although one-way functions or pre-image resistant hash functions
have a long history of being used for linking messages in message
authentication \cite{chanme03,merkle89}, it should be noted that the
direct application of a one-way function as the key generator is not
sufficient to achieve the goal of key secrecy here.\footnote{Recall
that a one-way function is one which is easy to evaluate in one
direction but hard in the reverse. In some implementation, the
output of a one-way function may leak a significant fraction of the
input bits. For example, suppose $f: \{0,1\}^l \rightarrow
\{0,1\}^l$ is a one-way function leaking no input bit, we could
construct another one-way function $f': \{0,1\}^{2l} \rightarrow
\{0,1\}^{2l}$ in the following way: $f'(x_1 || x_2) = x_1 || f(x_2)$
where $x_1, x_2 \in \{0,1\}^l$. This is still a one-way function but
leaks half of the input bits. Consequently, one might be able to
distinguish between its output and a uniformly picked random
number.} A more careful composition of one-way functions is needed
for linking keys together, namely, a pseudorandom function (PRF)
whose definition is as follows.

\begin{Def}
Let $f: \{0,1\}^{l_s} \times \{0,1\}^{l_i} \rightarrow
\{0,1\}^{l_o}$ be a function which takes a seed key $s \in
\{0,1\}^{l_s}$ and an input string $x \in \{0,1\}^{l_i}$ and outputs
another string $y \in \{0,1\}^{l_o}$ (i.e. $y = f_s(x)$).
$f_s(\cdot)$ is is said to be taken from a pseudorandom function
ensemble\footnote{We will call $f_s$ a pseudorandom function for the
sake of simplicity despite the loss of rigor.} with index $s$ if it
satisfies that, with $s$ uniformly picked from $\{0,1\}^{l_s}$ and
kept secret, all computationally efficient algorithms ${\cal A}$
given a set $Z = \{(x', y'): y' = f_s(x')\}$ of evaluations of $f_s$
at $x' \in X$ of his choice could tell whether a given $y$ is the
output of $f_s(\cdot)$ on input $x \not \in X$ or randomly picked
form $\{0,1\}^{l_o}$ with a negligible advantage in $l_s$ for all
$x$, where the advantage of an algorithm ${\cal A}$ for a given $x$
is defined as follows:
\[
\left |
\begin{array}{l}
Pr[s \leftarrow \{0,1\}^{l_s}; y = f_s(x): {\cal A} (Z, x, y) = 1] -
Pr[y \leftarrow \{0,1\}^{l_o}: {\cal A}(Z, x, y) = 1]
\end{array}
\right | .
\]
\end{Def}

Suppose $f$ is a PRF.  To generate a key $k'$ for a resource (or
privileged group) with label $r'$ from another key $k$ for a
resource with label $r$, we could consider the concatenation of the
labels $r||r'$ as an input string to $f$ and generate $k'$ as $k' =
f_{k}(r||r')$. In the next section, privileged group identities in a
sensor network are used as resource labels.  The property of $f$
ensures that nobody (computationally bounded), without the knowledge
of $k$, would be able to distinguish $k'$ from a key directly picked
from the key space with a non-negligible advantage. This also
guarantees that nobody could extract $k$ from $k'$. If there is a
PPT algorithm ${\cal A}$ which can extract $k$ from $k'$, then it
could be used to tell whether a given $k'$ is generated form $f$ or
randomly picked as follows: run ${\cal A}$ on $k'$ to extract $k$
and check if $k' \isequal f_{k}(r||r')$; $k'$ is a generated from
$f$ if and only if $k' = f_{k}(r||r')$; otherwise, flip a coin to
make a random/wild guess. Hence, the key extraction problem is at
least as difficult as solving the decisional problem non-negligibly
better than a wild guess.  Conjectured pseudorandom functions which
are efficient for the purpose here include AES-OMAC \cite{BCK96} and
SHA-HMAC \cite{IK03}.  For example, if $h(\cdot)$ denotes the HMAC
function, $f_k(x)$ can simply be implemented as $h(k||x)$ where
$k||x$ denotes the concatenation of the secret key $k$ and the
public input $x$.

It is natural to worry about whether such indistinguishability
preserves if $f$ is used to generate a series of keys, that is,
whether $k_t$ is still computationally indistinguishable from a
random key if it is generated from $k$ in the following series: $k_1
= f_{k}(r||r_1)$; $k_2 = f_{k_1}(r_1 || r_2)$; $\ldots$;
$k_t=f_{k_{t-1}}(r_{t-1}||r_t)$. The following theorem would be
useful in answering this question.

\medskip
\begin{Thm}\label{thm_prf}
Suppose $f: \{0,1\}^{l_k} \times \{0,1\}^* \rightarrow
\{0,1\}^{l_k}$ is a PRF, $k$ is uniformly picked from
$\{0,1\}^{l_k}$, and $k_1 = f_{k}(r||r_1)$; $k_2 = f_{k_1}(r_1 ||
r_2)$; $\ldots$; $k_t=f_{k_{t-1}}(r_{t-1}||r_t)$.  If $t$ is
polynomially many in $l_k$, then $\{k_t\} \cong U_{l_k}$ (denoting
the two distributions are computationally
indistinguishable\footnote{That is, no polynomial time algorithm can
distinguish whether a given sample is from the former or latter
distributions.}) where $\{k_t\}$ is the distribution of $k_t$ and
$U_{l_k}$ is the uniform distribution over $\{0,1\}^{l_k}$.
\end{Thm}
\begin{Proof}
Suppose we look at the generation of $k_i$ from $k_{i-1}$ and assume
that $\{k_{i-1}\} \cong U_{l_k}$ with an indistinguishability
coefficient $\epsilon_{i-1}$ (defined as the maximum
indistinguishability advantage achievable by any poly-time
algorithm); that is, $\epsilon_{i-1}$ is negligible.  We know that
$k_i = f_{k_{i-1}}(r_{i-1} || r_i)$ and wish to show that $\{k_i\}
\cong U_{l_k}$.  We use the standard hybrid argument with the hybrid
distribution $K_i' = \{k_i': s_i \leftarrow \{0,1\}^{l_k}; k_i' =
f_{s_i} (r_{i-1} || r_i) \}$.  From the property of the PRF, $K_i'
\cong U_{l_k}$ with an indistinguishability coefficient
$\epsilon_{f}$ (negligible). We argue that $K_i' \cong \{k_i\}$ by
contradiction. Suppose there is a PPT algorithm ${\cal A}$ which can
distinguish between $K_i'$ and $\{k_i\}$ with a distinguishability
advantage $\epsilon_i'$, then it can be used to distinguish between
$\{k_{i-1}\}$ and $U_{l_k}$.

The construction is as follows: for a given $s \in \{0,1\}^{l_k}$,
compute $k = f_{s}(r_{i-1}||r_i)$ and run ${\cal A}$ on $k$.  If $s
\in \{k_{i-1}\}$, then $k \in \{k_i\}$, whereas, if $s \in U_{l_k}$,
then $k \in K_i'$.  Thus this perfectly simulates the challenge of
${\cal A}$ in a real attack and could be used to distinguish between
$\{k_{i-1}\}$ and $U_{l_k}$ (a contradiction to our assumption).
Hence, $\epsilon_i' \le \epsilon_{i-1}$.

Overall, $\{k_i\} \cong U_{l_k}$ with an indistinguishability
coefficient $\epsilon_i \le \epsilon_i' + \epsilon_f =
\epsilon_{i-1} + \epsilon_f$.  Note that when $i=1$, $\epsilon_0 =
0$ since $k_0 = k \in U_{l_k}$.  Summing over $i$, we have
$\epsilon_t \le t \epsilon_f$.  Since $\epsilon_f$ is negligible in
$l_k$, if $t$ is polynomially many, then $\epsilon_t$ remains
negligible in $l_k$.  This concludes the proof.
\end{Proof}

Since the security guarantee of a pseudorandom function is
computationally complexity based, key linking based on a
pseudorandom function is computationally secure.

\section{\label{ks-sensor} Key Linking for Pairwise Key Pre-distribution in Sensor Networks}

In this section, we look at three examples of applying key linking
to sensor network key pre-distribution (KPS).  In a sensor network,
each node is preloaded with a set of keys in its key ring in such a
way that it can establish a pairwise key with another node in its
physical neighborhood with reasonably high probability (mainly for
mutual entity authentication); the model considered here is the same
as that in \cite{esch02,CPS03,CY04,DDHV03,DDHCV04}. In pairwise KPS,
each privileged group consists of two users or nodes. We will ignore
the repeated usage of keys, which trades off security for reduced
key storage; but the discussion below should also apply to that
case.

\subsection{A Graph-theoretic Representation of KPS for Sensor Networks}
When two sensor nodes share a common key, they can mutually
authenticate each other.  We can easily represent this keying or
trust relationship in a graph; that is, the sensor nodes are
represented as vertices and an edge exists between two vertices if
the corresponding sensor nodes share a common key.  This graph is
called a keying relationship graph in the following discussion. Note
that the keying relationship graph is a logical graph and does not
reflect the actual network topology of the sensor network during
deployment.  We assume there are $n$ sensors.

\subsection{Key Linking for KPS in Sensor Networks}
\paragraph{Example 1 --- KPS for sensor networks with one or multiple base stations \cite{SWCT01}}
In \cite{SWCT01}, the base station of a sensor network (with $n$
nodes) has a master key which is used (with PRF) to derive different
keys, with each one being shared between the base station and a
different sensor node. That is, each sensor node and the base
station only needs to store a single key. The keying relationship
graph is simply a star with the base station at the centre.  This is
indeed a special instance of the access structure graph discussed in
Section \ref{access}; here, the number of users is $(n+1)$
(including the base station) and the number of resources is $n$.
Applying Theorem \ref{ch2-thm-hash}, the maximum key storage in the
best case is $\lceil \frac{n}{n+1} \rceil = 1$.  Hence, the design
in \cite{SWCT01} is indeed optimal in its context.  By a similar
token, we could apply Theorem \ref{ch2-thm-hash} to cases with
multiple base stations.

\paragraph{Example 2 --- KPS with perfect connectivity in the key relationship graph}
Ideally, to ensure any pair of physical neighboring nodes in the
deployed network to be able to find a shared key, each node needs to
store $(n-1)$ keys (without key linking) if there are $n$ nodes
labeled from $0$ to $(n-1)$.  That is, the keying relationship graph
${\cal G}$ is a complete graph. This storage requirement is
trivially impractical. Since there are $\binom{n}{2}$ possible
groups, if key linking is applied, the maximum key storage in the
best case is $\lceil \binom{n}{2}/n \rceil = \lceil
\frac{n(n-1)}{2n} \rceil = \lceil \frac{n-1}{2} \rceil$ (Theorem
\ref{ch2-thm-hash});the maximum possible reduction factor is
$\frac{1}{2}$ (still not good enough).

The implementation of the linking could be done as follows. Without
loss of generality, assume $n$ is odd.  A user $i$ needs to store
one seed key $k_i$ and $\frac{n-1}{2}$ other derived keys $\{k_{ji}
= f_{k_j}(j||i): j = (i-d) \text{ mod } n, d \in [1, \frac{n-1}{2}]
\}$ where $k_{ji}$ is the pairwise key between node $j$ (where $j =
(i-d) \text{ mod } n, d \in [1, \frac{n-1}{2}]$) and node $i$.  For
the pairwise key between node $i$ and node $j'$ (where $j' = (i+ d)
\text{ mod } n, d \in [1, \frac{n-1}{2}]$), $k_{ij'} =
f_{k_i}(i||j')$.  That is, for the $\frac{n-1}{2}$ nodes in front of
node $i$, node $i$ has to store the derived keys, whereas, for the
$\frac{n-1}{2}$ nodes behind it, it can derive the pairwise key from
$k_i$.  As a result, the overall key storage per node is
$\frac{n-1}{2} + 1$.

\paragraph{Example 3 --- KPS with bounded connectivity in the key relationship graph}
In many cases, due to storage constraint, each node can only share a
common key with another $c$ nodes with $c < n$.  That is, each
vertex in the keying relationship graph has a bounded degree.
Without loss of generality, assume $c$ is even.

The total number of edges of the resulting keying relationship graph
${\cal G}'$ is $\frac{nc}{2}$ which is the total number of possible
groups. If key linking is applied, the maximum key storage per node
in the best case is $\lceil \frac{nc}{2n} \rceil = \lceil
\frac{c}{2}\rceil$; the maximum reduction factor is again
$\frac{1}{2}$.  In the best possible case, a node $i$ would only
need to store one key $k_i$ and $\frac{c}{2}$ other derived keys
$k_{ji}$.  The problem of determining which half of the $c$ pairwise
keys are derived from $k_i$ and which half are obtained from other
$\frac{c}{2}$ nodes could be solved by finding an Eulerian tour over
${\cal G}'$.  An Eulerian tour over a graph $G$ is a tour along the
edges of $G$ so that each edge is passed exactly once.  Such a tour
exists in a graph $G$ if $G$ has at most two vertices with an odd
degree; this is fulfilled for ${\cal G}'$ in question. The Fleury's
algorithm (shown in Appendix) with running time $O(|E(G)|)$ (where
$|E(G)|$ is the total number of edges in $G$) can be used for
finding an Eulerian tour \cite{CLRS01} and the set of edges of each
vertex would be partitioned by the tour into two halves, one marked
as incoming edges and the other as outgoing edges.  Now we could
derive all pairwise keys on an outgoing edge of node $i$ using $k_i$
and set the keys of the incoming edges as $k_{ji}$ derived from
$k_j$ of another node $j$.

Regarding the case with an odd number of edges in a connectivity
graph, edges could be added to make the graph Eulerian.  If there
are vertices in a connectivity graph $G$ with an odd number of
edges, the total number of such vertices should be
even.\footnote{The sum of the degrees of all vertices of a graph is
even.  If a vertex has an odd degree, then there must exist another
vertex with an odd degree to make the total sum even.  That is,
vertices with an odd degree come in pairs.}  We could simply
partition such a subset of vertices into pairs and assign an edge to
each pair, then the resulting graph is Eulerian.

While a regular keying relationship graph is considered in this
example, the result and technique apply to a more general keying
relationship graph as long as one can partition the edges connecting
each vertex into two parts.  Reduction on maximum key storage could
always be achieved.

Recall that given any key of a node on a particular {\em outgoing}
edge of the key relationship graph, it is computationally infeasible
to find the keys on other outgoing edges of the node, thus
guaranteeing the resilience of compromised nodes.  Any collusion of
compromised nodes would not threat the security of the remaining
nodes since we have considered an ideal access structure and it is
computationally hard for the collusion to find any key not
originally held by them if a pseudorandom function is used.

\section{\label{discuss} Discussions}
While there is always reduction in the average storage whenever
there is redundant membership, key linking may not lead to reduction
on the maximum key storage per user in some access structure.  Under
the framework of constraints considered in this paper, it could be
difficult to achieve reduction on the maximum storage in those
cases.  An analogy to this situation is when Huffman encoding for
the equiprobable case.  This sets the limits of any scheme if
trading off security is not considered. If further reduction on the
maximum key storage is necessary, trading off the ideal access
structure is one possible solution and combinatorial techniques
could apply as in \cite{blun92,BC96,mitc88,dyer95}. Alternatively,
we could consider a set of keys as a long bit string (instead of a
set of individual keys) and create the linking on a bit-by-bit basis
using the technique of correlated pseudorandomness \cite{CDI05};
however, the gain also comes as a result of trading off security;
now a non-privileged user may learn some of the bits of a resource
key he is not supposed to whereas the key linking technique
considered in this paper would not leak out information that can be
efficiently extracted by a non-privileged user.

\section{\label{conclude} Conclusion}

As applications involving low cost devices like sensor networks and
RFID emerge, memory cost (for secure key storage) which is usually
not a concern has become an essential constraint to designing
security systems.  To alleviate this, the key storage requirement
could possibly reduced by creating dependency among secret keys
stored in a user device, that is, key linking.  Key linking exploits
redundancy in privileged group memberships for key compression.

We derive an upper bound for maximum achievable key compression in a
system with ideal, general access structure.  This bound is tight
and can somehow be treated as a negative result, which demonstrates
that without trading off security, considerable key storage
reduction may not be achievable. We also show a provably secure
instantiation of key linking scheme using pesudorandom functions. We
show how to apply the key linking technique to reduce key storage in
pairwise key pre-distribution in wireless sensor networks; the
storage reduction is still not sufficient to give efficient schemes
which again demonstrate the cost in efficiency loss we have to pay
if no security tradeoff is considered.  The results actually provide
ground for proposals which trade off security for efficiency.

\bibliographystyle{plain}
\bibliography{./thesis,./mybook,./crypto,./adhocsec,./key,./trc,./cda}

\appendix{\centerline{\bf Fleury's Algorithm}}

Fleury's algorithm constructs an Euler circuit in a graph (if it's
possible).  The algorithm runs as follows:
\begin{enumerate}
\item Pick any vertex to start.

\item From that vertex pick an edge to traverse, considering
following rule: never cross a bridge of the reduced
graph\footnote{By "reduced graph" we mean the original graph minus
the darkened (already used) edge.} unless there is no other choice.

\item Darken that edge, as a reminder that you can't traverse it
again.

\item Travel that edge, coming to the next vertex.

\item Repeat Steps 2-4 until all edges have been traversed, and you are
back at the starting vertex.
\end{enumerate}

\end{document}